# Quantum Field Theoretical Treatment of Neutrino Oscillations

Viliam Pažma[1], Július Vanko[2] and Juraj Chovan[2]

[1] Institute of Physics, Comenius University, Bratislava, Slovakia
[2] Department of Nuclear Physics, Comenius University, Bratislava, Slovakia

vanko@fmph.uniba.sk

The quantum field theoretical treatment of neutrino oscillations is formulated. The procedure how to calculate the transition probability of one type neutrino into another one is briefly outlined.

1. Introduction

The field theoretical treatments of neutrino oscillations were studied and discussed in many papers (see e.g. [1-8] and references therein). In spite if this one can hardly say that all conceptual questions (many of them are mentioned and discussed, e.g., in [1]) are resolved. For example, in [2] argument were given that a Fock space of (flavor) neutrinos does not exists. It would be desirable to understand to a what extent are those arguments based on hypothesises of the standard theory of neutrino oscillations or whether are they valid in general (i.e. do they represent some physical principles) ?
    We have no ambitious to answer these and similar question fully satisfactorily. We want, simply speaking, offer a new look on those problems and we hope that it can enlight some of above mentioned problems.
    The theory of neutrino oscillations was formulated to solve the solar neutrino puzzle mainly. It seems to be natural to expect that such theory must contain something specifying a source in an explicite form. In the framework of the standard theory of neutrino oscillations a source does not play, in general, any important role. Can this fact cause some of known conceptual problems ?
    Trying to answer this question we, in our previous paper [9], proposed the alternative theory of neutrino oscillations. Because neutrino oscillations are hard studied mainly in solar neutrino experiments [10] we proposed the equation

$$i\partial_t \Psi - H\Psi = i\varphi_0(\vec{x},t) \qquad (1)$$

as the equation describing neutrinos (and their oscillations) produced by a source (e.g. Sun). (Several reasons leading to (1) can be found in [9] ).

The hamiltonian $H$ in (1) is given by

$$H = \vec{\alpha}.\vec{p} + \beta M = H_0 + \beta M' ,\qquad(2)$$

where $(M')^+ = M'$, $M_{ii}' = 0$, for $i = 1, 2, 3$, $M_d = M - M' = diag(m_1, m_2, m_3)$, $m_i$ are masses of
$\nu_i$ ($\nu_1 = \nu_e$, $\nu_2 = \nu_\mu$, $\nu_3 = \nu_\tau$) and standard meaning of other symbols is assumed.

The free $\nu_i$ is described (in p-representation) by the amplitude $U_i$ satisfying

$$\varepsilon_i U_i = (\vec{\alpha}.\vec{p} + \beta M_d) U_i , \quad \left(\varepsilon_i = \sqrt{\vec{p}^2 + m_i^2}\right), \quad \bar{U}_i U_i = 2m_i ,\qquad(3)$$

and corresponding to the negative helicity.

If a source produces, e.g., $\nu_1$ neutrinos only then for $\varphi_0$ we proposed the following expression

$$\varphi_0(\vec{x},t) \sim \int d^3\vec{x}_0 \, \rho_s(\vec{x}_0,t) \, N(\vec{x}_0,t) \int \frac{d^3\vec{p}}{(2\pi)^{\frac{3}{2}}} \, C(\vec{p},\vec{x}_0,t) \, \varepsilon_1 \frac{U_1}{\sqrt{2\varepsilon_1}} \, e^{i\vec{p}\cdot(\vec{x}-\vec{x}_0)} ,\qquad(4)$$

where $\rho_s(\vec{x}_0,t)$ is the density of point sources (in this case we define any source as a set of point sources), $N(\vec{x}_0,t)$ is the number of neutrinos produced by point source (placed at $\vec{x}_0$) per unit of time, $C(\vec{p},\vec{x}_0,t)$ describes the spectrum of neutrinos produced by that point source (more details see in [9]).

The equation (1) can be written in covariant form as

$$(i\gamma^\mu \partial_\mu - M)\Psi = i\gamma^\mu \varphi_\mu ,\qquad(5)$$

where

$$\varphi_\mu(x) \sim \int d^4x_0 \, J_\mu(x_0) N(x_0) \int d^4p \, \tilde{C}(p,x_0) \, v.p \, \Theta(p_0) \, \delta(p^2 - m^2) \, \delta(v.(x-x_0)) \, U_1 e^{-ip\cdot(x-x_0)} ,$$

$$(6)$$

and

$$J_\mu = \left( \frac{\rho_s}{\sqrt{1-v^2}}, \frac{-\vec{v}\rho_s}{\sqrt{1-v^2}} \right),$$

$$v_\mu = \left( \frac{1}{\sqrt{1-v^2}}, \frac{-\vec{v}}{\sqrt{1-v^2}} \right)$$

and $\vec{v}$ is the velocity of any part of considered source with respect to an observer.

We do not write down explicitly a constant of the proportionality in (4) or (5) because that constant will play no role in the next considerations.

2. Quantum field theory of neutrino oscillations

The lagrangian corresponding to (3) can be written in the form

$$L = \int d^3\vec{x} \left[ \overline{\Psi}(i\gamma^\mu \partial_\mu - M)\Psi - i\left( \overline{\Psi}\gamma^\mu \varphi_\mu - \overline{\varphi}_\mu \gamma^\mu \Psi \right) \right]. \tag{7}$$

The hamiltonian operator $\widehat{H}$ of the considered system is

$$\widehat{H} = \int d^3\vec{x} \left[ \widehat{\overline{\Psi}}\left( \vec{\alpha}.\widehat{\vec{p}} + \beta M \right)\widehat{\Psi} + i\left( \widehat{\overline{\Psi}}\gamma^\mu \varphi_\mu - \overline{\varphi}_\mu \gamma^\mu \widehat{\Psi} \right) \right] = \widehat{H}_0 + \widehat{H}\prime, \tag{8}$$

where

$$\widehat{H}_0 = \int d^3\vec{x}\, \widehat{\overline{\Psi}}\left( \vec{\alpha}.\widehat{\vec{p}} + \beta M_d \right)\widehat{\Psi},$$

$$\widehat{H}\prime = \int d^3\vec{x} \left[ \widehat{\overline{\Psi}}\beta M\prime\widehat{\Psi} + i\left( \widehat{\overline{\Psi}}\gamma^\mu \varphi_\mu - \overline{\varphi}_\mu \gamma^\mu \widehat{\Psi} \right) \right]$$

and $\widehat{\Psi}(x)$ is the neutrino field operator.

In the interaction representation (up to here we have been considering $\widehat{H}\prime$ as the hamiltonian of interaction and in the interaction representation we shall denote it as $\widehat{H}_{int}$) the field operator $\widehat{\Psi}(x)$ satisfies the equation for free field and thus we can write

$$\widehat{\Psi}(x) = \sum_{k=1}^{3}\sum_{\sigma}\int \frac{d^3\vec{p}}{(2\pi)^{\frac{3}{2}}}\left[a_k(\vec{p},\sigma)\frac{U_k(p_k,\sigma)}{\sqrt{2\varepsilon_k}}e^{-ip_kx} + b_k^+(-\vec{p},-\sigma)\frac{U_k(-p_k,-\sigma)}{\sqrt{2\varepsilon_k}}e^{ip_kx}\right],$$

where $p_k = (\varepsilon_k, \vec{p}) = (p_k^\mu)$, $U_k(p_k,\sigma)$ are solutions to (3) corresponding to helicity $\sigma$ ($\sigma = \pm 1$) and $a_k$, $a_k^+$, $b_k$, $b_k^+$ are governed by standard anticommuting relations.

Now the time development of states $\Phi$'s is determined by the equation

$$i\partial_t \Phi = \widehat{H}_{int}\Phi. \tag{9}$$

3. Application

Let us now consider a stationary source ($\rho_s$, $N$, $C$ are independent on time and moreover $C = C(\vec{p})$, i.e. a considered source is composed of the identical point sources) which does not moves with respect to an observer (i.e. $\vec{v} = 0$). Thus we deal with $\varphi_\mu$ which is of the form $\varphi_\mu = 0$ ($\mu$=1, 2, 3) and

$$\varphi_0 = \int d^3\vec{x}_0 \, \rho_s(\vec{x}_0) N(\vec{x}_0) \int \frac{d^3\vec{p}}{(2\pi)^{\frac{3}{2}}} C(\vec{p}) \varepsilon_1 \frac{U_1}{\sqrt{2\varepsilon_1}} e^{i\vec{p}\cdot(\vec{x}-\vec{x}_0)} =$$

$$= \int \frac{d^3\vec{p}}{(2\pi)^{\frac{3}{2}}} F(\vec{p}) C_1(\vec{p}) \frac{U_1}{\sqrt{2\varepsilon_1}} e^{i\vec{p}\cdot\vec{x}}. \tag{10}$$

Let us now consider as the initial state (in t=-∞) the vacuum state $|0\rangle$. This state will can develope into the state

$$|1,\vec{p}\rangle = (2\pi)^{\frac{3}{2}} a_k^+(\vec{p},\sigma = -1)|0\rangle$$

in time t with the amplitude of probability $A(\nu_1 \to \nu_k; t)$. In the lowest order of the perturbative theory this amplitude is equal to

$$-\langle k,\vec{p}|\int_{-\infty}^{t} dt_2 \widehat{H}_{2int} \int_{-\infty}^{t_2} dt_1 \widehat{H}_{1int}|0\rangle,$$

where

$$\widehat{H}_{1int} = i\int d^3\vec{x}\, \left(\widehat{\overline{\Psi}}\gamma^0\varphi_0 - \overline{\varphi}_0\gamma_0\widehat{\Psi}\right),$$

$$\widehat{H}_{2int} = \int d^3\vec{x}\, \widehat{\overline{\Psi}}\gamma^0 M'\, \widehat{\Psi}.$$

Thus we get ($k=2, 3$)

$$A(v_1 \to v_k; t) = i\frac{\overline{U}_k(p_k)M'_{k1}U_1(p_k)}{2\sqrt{\varepsilon_k\varepsilon_1}} \cdot \frac{F(\vec{p})C_1(\vec{p})}{\varepsilon_k\varepsilon_1}\, e^{i\varepsilon_k t} = A_{k1}(\vec{p})\, e^{i\varepsilon_k t}. \qquad (11)$$

Trying calculate the flux of neutrinos $v_k$ (having certain momenta) at some point $\vec{x}$ in time $t$ we have to have at the elbow the wave function $\Psi$ in the x-representation. This wave function must correspond to the state

$$|v\rangle = -\int_{-\infty}^{t} dt_2\, \widehat{H}_{2int} \int_{-\infty}^{t_2} dt_1\, \widehat{H}_{1int}|0\rangle.$$

The standard prescription for sought $\Psi(x)$ is

$$\Psi(x) = \langle 0|\widehat{\Psi}(x)|v\rangle = \sum_{k,\vec{p},\sigma}\langle 0|\widehat{\Psi}(x)|k,\vec{p},\sigma\rangle\langle k,\vec{p},\sigma|v\rangle.$$

Thus we get

$$\Psi(x) = \sum_{k=2}^{3}\int \frac{d^3\vec{p}}{(2\pi)^{\frac{3}{2}}} \frac{U_k}{\sqrt{2\varepsilon_k}}\, e^{-ip_k x}\, A(v_1 \to v_k; t) =$$

$$= \sum_{k=2}^{3}\int \frac{d^3\vec{p}}{(2\pi)^{\frac{3}{2}}}\, A_{k1}(\vec{p})\, \frac{U_k}{\sqrt{2\varepsilon_k}}\, e^{i\vec{p}\cdot\vec{x}} \equiv \Psi(\vec{x}). \qquad (12)$$

Defining $\Psi_k^{(\vec{n})}(\vec{x})$ by means of

$$\Psi_k(\vec{x}) = \int \frac{d^3\vec{p}}{(2\pi)^{\frac{3}{2}}} A_{k1}(\vec{p}) \frac{U_k}{\sqrt{2\varepsilon_k}} e^{i\vec{p}\cdot\vec{x}} =$$

$$= \int d\Omega(\vec{n}) \int_0^\infty \frac{dp\, p^2}{(2\pi)^{\frac{3}{2}}} A_{k1}(p\vec{n}) \frac{U_k}{\sqrt{2\varepsilon_k}} e^{ip\vec{n}\cdot\vec{x}} = \int d\Omega(\vec{n}) \, \Psi_k^{(\vec{n})}(\vec{x}) \, .$$

then the density of the flux $\vec{j}_k^{(\vec{n})}(\vec{x})$ of neutrinos having momenta $\vec{p} = p\vec{n}$ ($p \in (0,\infty)$) at the point $\vec{x} = \vec{n}L$ is

$$\vec{j}_k^{(\vec{n})}(\vec{n}L) = \Psi_k^{+(\vec{n})}(\vec{n}L) \, \vec{\alpha} \, \Psi_k^{(\vec{n})}(\vec{n}L) \, .$$

Now the probability $P(\nu_1 \to \nu_k; \vec{n}L)$ of the transition $\nu_1 \to \nu_k$ can be defined as

$$P(\nu_1 \to \nu_k; \vec{n}L) = \frac{\left|\vec{j}_k^{(\vec{n})}(\vec{n}L)\right|}{\left|\vec{j}_{tot}^{(\vec{n})}(\vec{n}L)\right|} \, .$$

The quantity $\vec{j}_{tot}^{(\vec{n})}$ we can determine from the wave function of produced neutrinos if we ignore the oscillations (i.e. if we put $M' = 0$).

4. Conclusion

The presented considerations are applications of standard procedures to certain systems only. These systems may seem to be unusual but they are physically real and we, simply speaking, try to describe them. In our opinion the results following from from this briefly outlined considerations seem to be acceptable at first glance at least. However, next more detailed investigations are necessary. Namely, we are not assured whether a naive use of the perturbative theory is admissible. Moreover we feel that some problems can rise as to the choice of the initial state (in this paper we choose $|0\rangle$ as the initial state).

Another problems can also rise as to the interpretation because, as far as we know, systems of the considered type were not studied up to now.